\pdfoutput=1
\documentclass{iau}
\usepackage{graphicx}
\usepackage{pifont}
\usepackage{natbib}
\newcommand{\bm}{\boldmath}
\newcommand{\aap}{A\&~A}
\newcommand{\mnras}{Mon.~Not.~Roy.~Soc.}
\newcommand{\apj}{ApJ}
\newcommand{\apjl}{ApJL}
\newcommand{\physrep}{Phys.~Report}
\newcommand{\pre}{Phys.~Rev.~E}
\newcommand{\solphys}{Sol.Phys.}

\usepackage{setspace}

\newcommand\epigraph[3]{
\vspace{1em}\hfill{}\begin{minipage}{#1}{\begin{spacing}{0.9}
\small\noindent\textit{#2}\end{spacing}
\vspace{1em}
\hfill{}{#3}}\vspace{2em}
\end{minipage}}

\usepackage{amsmath}
\usepackage{amssymb}
\bibpunct{(}{)}{;}{a}{}{,}

\title[] 
{Advances in mean-field dynamo theories}

\author{V.V. Pipin$^1$}

\affiliation{$^1$Institute of Solar-Terrestrial Physics, Russian Academy of
Sciences, \\ Irkutsk, 664033, Russia,  email: {\tt pip@iszf.irk.ru} }

\pubyear{2013}
\volume{294}  

\pagerange{--}

\jname{Solar and Astrophysical Dynamos and Magnetic
Activity}
\editors{A.G. Kosovichev, E.M. de Gouveia Dal Pino, \& Y.Yan, eds.}

\begin{document}
\maketitle
\epigraph{3.in}{
Is dynamo in tachocline or in convection zone? \\
That is the question! \\
Whether 'tis nobler in the mind to fit \\
Parameters of model to reproduce the observations \\
Or to take  the theory and direct simulations \\
And try to answer basic questions: \\
Why magnetic field is generated? \\
And why it drifts equatorward in course of solar cycle?\\
'tis a consummation devotely to be wished!}{take-off  Shakespeare}

\begin{abstract}
We give a short introduction to the subject and review advances in
understanding the basic ingredients of the mean-field dynamo theory.
The discussion includes the recent analytic and numerical work in
developments for the mean electromotive force of the turbulent flows
and magnetic field, the nonlinear effects of the magnetic helicity,
the non-local generation effects in the dynamo. We give an example
of the mean-field solar dynamo model that incorporates the fairly
complete expressions for the mean-electromotive force, the subsurface
shear layer and the conservation of the total helicity. The model
is used to shed light on the issues in the solar dynamo and on the
future development of this field of research.
\end{abstract}

\section{Introduction}

The mean-field magnetohydrodynamic presents one of the most powerful
tools for exploring the nature of the large-scale magnetic activity
in cosmic bodies \citep{moff:78,park,krarad80}. It is widely believed
that magnetic field generation there is governed by interplay between
turbulent motions of electrically conductive fluids and global rotation.
The dynamo theory studies the evolution of the magnetic field which
is govern by the induction equation:
\[
\frac{\partial\mathbf{B}}{\partial t}=\nabla\times\left(\mathbf{U\times B}-\eta\nabla\times\mathbf{B}\right),
\]
where $\mathbf{B}$ is the magnetic field induction vector, $\mathbf{U}$
is the velocity field of the plasma and $\eta$ is the molecular magnetic
diffusivity.

\section{The mean electromotive force}

The aim of this section is to briefly outline the basic equations
and methods of the mean-field magnetohydrodynamic (MHD). The general
framework of the mean-field MHD can be introduced as follows. In the
turbulent media, it is feasible to decompose the fields on the mean
and fluctuated parts, e.g., $\mathbf{B}=\overline{\mathbf{B}}+\mathbf{b},\,\mathbf{U}=\mathbf{\overline{\mathbf{U}}}+\mathbf{u}$.
Hereafter, everywhere, we use the small letters for the fluctuating
part of the fields and capital letters with a bar above for the mean
fields. Let's define the typical spatial, $L,\ell$ , and temporal,
$T,\tau_{c}$, variation scales for the mean and the fluctuated parts
of the fields. It is a typical situation for the astrophysical system
when the flows and magnetic fields are strongly turbulent, i.e., the
Strouhal number $St={\displaystyle \frac{u\tau_{c}}{\ell}\sim1}$
and the Reynolds number $R_{e}={\displaystyle \frac{u\ell}{\nu}}\gg1$,
where $\nu$ is the molecular viscosity. Assuming the validity of
the Reynolds rules \citep{1975mit..bookR....M} and averaging the
induction equation over the ensemble of the fluctuating fields we
get the mean-field dynamo equation: 
\begin{equation}
\frac{\partial\overline{\mathbf{B}}}{\partial t}=\nabla\times\left(\mathbf{\overline{U}\times}\overline{\mathbf{B}}+\boldsymbol{\mathcal{E}}\right),\,\,\,\boldsymbol{\mathcal{E}}\mathbf{=}\overline{\mathbf{u}\times\mathbf{b}}\label{eq:mf}
\end{equation}
There are two contributions here. In a perfectly conducting fluid,
when the magnetic Reynolds number $R_{m}={\displaystyle \frac{u\ell}{\eta}}\gg1$,
the magnetic flux is frozen into fluid. Then, the first term can be
interpreted as the defection, stretching and compression (or expansion)
of the magnetic field by mean flow, because of $\mathbf{\nabla\times\left(\overline{U}\times\overline{\mathbf{B}}\right)}=-\left(\overline{\mathbf{U}}\cdot\nabla\right)\mathbf{\overline{B}}+\left(\mathbf{\overline{B}}\cdot\nabla\right)\mathbf{\overline{U}}-\overline{\mathbf{B}}\left(\nabla\cdot\overline{\mathbf{U}}\right)$.
The effect of the turbulence is represented by the mean electromotive
force $\boldsymbol{\mathcal{E}}\mathbf{=}\overline{\mathbf{u}\times\mathbf{b}}$.
It is possible to analyze the general structure of the $\boldsymbol{\mathcal{E}}$
using the assumption about the scale separation in the turbulence
$L,T\gg\ell,\tau_{c}$ and the transformation symmetry properties
of the basic physical quantities \citep{rad69,krarad80,bran2012AA}:

\begin{equation}
\boldsymbol{\mathcal{E}}=\left(\hat{\mathbf{\alpha}}+\hat{\mathbf{\gamma}}\right)\circ\overline{\mathbf{B}}-\hat{\eta}\circ\left(\nabla\times\overline{\mathbf{B}}\right)+\hat{\kappa}\circ\left(\boldsymbol{\nabla}\overline{\mathbf{B}}\right)+o\left(\frac{\ell}{L}\right),\label{eq:emf}
\end{equation}
where $\left(\boldsymbol{\nabla}\overline{\mathbf{B}}\right)_{\{i,j\}}=\frac{1}{2}\left(\nabla_{i}\overline{B}_{j}+\nabla_{j}\overline{B}_{i}\right)$,
the kinetic coefficients $\hat{\mathbf{\alpha}},\hat{\mathbf{\gamma}},\hat{\eta},\hat{\kappa}$
are tensors and the symbol $\circ$ marks the tensor product. 

The kinetic coefficients may depend on the global factors, which determine
the large-scale properties of the astrophysical system, for example,
the global rotation angular velocity $\boldsymbol{\Omega}$, the large-scale
vorticity $\mathbf{W}=\boldsymbol{\nabla}\times\overline{\mathbf{U}}$,
the stratification parameters like $\boldsymbol{\nabla}\overline{\rho}$,
$\boldsymbol{\nabla}\overline{\mathbf{u}^{2}}$, and the global constraints,
like, magnetic helicity conservation. For the simplest case when $\hat{\alpha}\sim\alpha_{0}\delta_{ij}$,
$\hat{\eta}\sim\eta_{T}\delta_{ij}$ and $\hat{\gamma}\sim-\varepsilon_{ijn}\overline{V}_{n}^{(p)}$
and $\hat{\kappa}$ is neglected we get\citep{krarad80}:
\begin{equation}
\mathcal{E}_{i}=\alpha_{0}B_{i}+\left(\mathbf{\overline{V}}^{(p)}\times\overline{\mathbf{B}}\right)-\eta_{T}\left(\nabla\times\overline{\mathbf{B}}\right),\label{eq:emfs}
\end{equation}
where $\alpha_{0}$ is the magnitude of the $\alpha$ effect, $\overline{\mathbf{V}}^{(p)}$
is the turbulent pumping velocity and $\eta_{T}$ is the isotropic
turbulent diffusivity. 

To calculate the kinetic coefficients we use the equations which govern
the evolution of the fluctuating magnetic and velocity fields. For
example, taking into account the effects of the global rotation and
shear for the incompressible turbulent flows, we get the equations
as follows (see, e.g., \citealp{rad-kle-rog,2005PhR...417....1B})
\begin{eqnarray}
\frac{\partial\mathbf{b}}{\partial t} & = & \nabla\times\left(\mathbf{u}\times\mathbf{\overline{B}}+\mathbf{\overline{U}}\times\mathbf{b}\right)+\eta\nabla^{2}\mathbf{b}+\mathfrak{G},\label{induc1-1}\\
\frac{\partial u_{i}}{\partial t}+2\left(\mathbf{\Omega}\times\mathbf{u}\right)_{i} & = & -\nabla_{i}\left(p+\frac{\left(\mathbf{b\cdot}\overline{\mathbf{U}}\right)}{\mu}\right)+\nu\Delta u_{i}\label{navie1-1}\\
 & + & \frac{1}{\mu}\nabla_{j}\left(\overline{B}_{j}b_{i}+\overline{B}_{i}b_{j}\right)-\nabla_{j}\left(\overline{U}_{j}u_{i}+\overline{U}_{i}u_{j}\right)+f_{i}+\mathfrak{F}_{i},\nonumber 
\end{eqnarray}
where $\mathfrak{G},\mathfrak{F}$ stand for the nonlinear contributions
of fluctuating fields, $p$ is the fluctuating pressure, $\mathbf{\Omega}$
is the angular velocity responsible for the Coriolis force, $\mathbf{f}$
is the random force driving the turbulence. 

Using Eqs(\ref{induc1-1},\ref{navie1-1}), $\boldsymbol{\mathcal{E}}$
can be calculated analytically by different ways. The first-order
smoothing, also known as the second order correlation approximation
(SOCA) uses the condition $\min\left(Rm,St\right)\ll1$ and neglect
the nonlinear contributions in Eqs(\ref{induc1-1},\ref{navie1-1})
\citep{moff:78,krarad80}. The $\tau$-approximations was introduced
to take into account the nonlinear effects of the second order correlations.
It is claimed to be valid for $Rm,Re\gg1$ and for the developed turbulence
in equilibrium state. In this case we solve the equations for the
second order correlations and replace the third-order correlations
of the fluctuating parameters, with the second order relaxation terms
(see details in \citealt{1996A&A...307..293K,2002PhRvL..89z5007B,2003GApFD..97..249R,2005PhR...417....1B}).
This approximation is based on questionable assumptions ( \citealt{2007GApFD.101..117R}),
e.g., it is assumed that the second-order correlations do not vary
significantly on the time scale of $\tau_{c}$. This assumption is
consistent with scale separation between the mean and fluctuating
quantities in the mean-field magnetohydrodynamic. The reader can
find a comprehensive discussion of the $\tau$ -approximation in the
above cited papers. 

The path integral approach use the ideas from the stochastic calculus
\citep{1984AN....305..119D,1984JFM...144....1Z}. This approach is
valid for the case $Re\gg1$, $Rm\ll1$. The reader can find the relevant
examples in papers by \citet{kle-rog99} and by \citet{garr2011}. 
The every analytic al method calculation of $\boldsymbol{\mathcal{E}}$
has to use the assumptions about the background turbulence which would
exist in the absence of the large-scale magnetic fields and flows
(e.g., global rotation and shear). For the numerical solution it is
equivalent to definition of the stochastic driving force $\mathbf{f}$.
Despite the Eqs(\ref{induc1-1},\ref{navie1-1}) is widely applied
to the dynamo in the Sun and the late-type stars, these equations
describe the forced isothermal turbulence rather than turbulent convection.
The latter also can be treated analytically using the $\tau$-approximation
\citep{2003PhRvE..67b6321K}. 

The mean-electromotive force can be estimated by the direct numerical
solution (DNS) of the equations like Eqs(\ref{induc1-1},\ref{navie1-1})
(e.g., \citealt{2001ApJ...550..824B,2007AN....328.1006K,2008ApJ...676..740B,2010PhFl...22c7101L,2011ApJ...727..127T})
or the similar ones for the turbulent convection by the so-called
``impose-field'' method (e.g., \citealt{oss2001,oss02}) or the
so-called ``test-field'' method (\citealt{2005AN....326..245S,2008A&A...491..353K,2009A&A...500..633K,2010A&A...520A..28R,2011A&A...533A.108S,2012ApJ...752..121S}).
The global simulations of the geo- and stellar dynamos also can be
used to extract the mean-field dynamo coefficients from simulations
(see the above cited papers and \citealt{2011ApJ...735...46R,2011ASPC..448..277B}).

\section{Mean-field phenomena in the solar magnetic activity}

There is a wide range of the magnetic activity phenomena which are
observed on the Sun and the others astrophysical systems. Here, I
restrict myself with consideration to the solar magnetic activity.
The observation of the solar magnetic fields shows that for the spatial
scales the basic assumption behind the Eqs(\ref{eq:mf},\ref{eq:emf})
is not fulfilled (see, e.g., the review by Sami Solanki in this volume).
The given theory is not able to capture self-consistently and simultaneously
the origin of the large-scale sunspots butterfly diagrams and the
emergence of the separate sunspots. Though the both phenomena can
be analyzed separately using the mean-field MHD framework, see, e.g.,
the application of the theory to the sunspot decay problem\citep{2000AN....321...75R}.
It is known that the large-scale temporal-spatial (e.g., time-latitude)
patterns, such as the sunspots butterfly diagrams, can be detected
for the much smaller scales phenomena, like ephemere regions\citep{1996SoPh..163..267M,2000eaa..bookE2275H,2004IAUS..223...49M}.
Therefore, even the scale-separation assumption is not valid for the
solar conditions we can consider the large-scale organization of the
magnetic activity phenomena as a manifestation of the large-scale
magnetic fields generated somewhere in the deep convection zone. This
idea is commonly adopted in the mean-field dynamo theory. 

The dynamo theory isolates of the details of processes, which are
responsible for the emergence of the magnetic activity features at
the surface, and studies the evolution of the large-scale magnetic
field govern by the dynamo equations Eqs(\ref{eq:mf},\ref{eq:emf}).
It is suggested that the toroidal part of the large-scale field forms
sunspots and organize the magnetic phenomena inside the Sun. The large-scale
poloidal field goes out of the Sun and governs the solar corona. Thus,
the key questions for the theory are the origin of the large-scale
magnetic activity spatial-temporal patterns, the phase relation between
activity of the poloidal and toroidal components, what defines the
solar cycle period and magnitude etc. Another portion of the problems
which could be studied using the same framework is related to the
statistical properties of the large-scale spatial organization of
the small-scale magnetic fields and motions in the solar convection
zone with the operating dynamo (see, e.g, \citealt{parn09,2012ApJ...745..129S}
and review by Jan Stenflo in this proceedings). 

The basic idea for the solar dynamo action was developed by \citet{par55}.
He suggested that the poloidal field of the Sun is stretched to the
toroidal component by the differential rotation ($\Omega$ effect)
and the cyclonic motions ($\alpha$ effect) return the part of the
toroidal magnetic field energy back to the poloidal component. This
is the so-called $\alpha\Omega$ scenario. While the turbulent diffusion
is not presented in the title, it is equally important \citep{park}.
The mean electromotive force given by Eq(\ref{eq:emfs}) fits to this
scenario. The $\Omega$ effect can be well understood because the
helioseismology provide the data about the distribution of the rotation
inside the convection zone and beneath. The big uncertainty is about
how the poloidal field of the Sun is generated. There is an ongoing
debate about a number of problems connected with the $\alpha$ effect
and $\alpha\Omega$ dynamos (see, e.g., \citealt{ruehol2004,2005PhR...417....1B}).
For instance, the period of the solar activity cycle poses a problem.
Namely, for mixing-length estimates of the turbulent magnetic diffusivity
in the convection zone and dynamo action distributed over the whole
convection zone, the obtained cycle periods are generally much shorter
than the observed 22 yr period of the activity cycle. For thin-layer
dynamos, the situation becomes even worse. 

The expression for the mean electromotive force $\boldsymbol{\mathcal{E}}$
contains a number of dynamo effects that may complement the $\alpha$
effect or may be an alternative to it. These effects are due to a
large-scale current, global rotation and (or) the large-scale shear
flow. The given dynamo effects are usually associated with the $\boldsymbol{\Omega}\times\mathbf{\overline{J}}$
effect \citep{rad69} and the shear-current effect $\overline{\mathbf{W}}\times\mathbf{\overline{J}}$
where, $\overline{\mathbf{W}}=\boldsymbol{\nabla}\times\overline{\mathbf{U}}$
\citep{kle-rog:04a}. In fact, these effects contribute to the antisymmetric
parts of $\hat{\eta}$ in the $\boldsymbol{\mathcal{E}}$ given by
Eq(\ref{eq:emf}). The reader can find the explicit expressions for
them in \citep{rad-kle-rog,kle-rog:04a,pi08Gafd}. \citet{2009A&A...493..819P}
and \citet{pk11mf} found that the inclusion of the additional turbulent
induction effects increases the period of the dynamo and brings the
large-scale toroidal field closer to the equator, thus improving the
agreement of the models with the observations. Also, in the models
the large-scale current dynamo effect produces less overlapping cycles
than dynamo models with $\alpha$ effect alone. The symmetric part
of $\hat{\kappa}$ (see, Eq. \ref{eq:emf}) contribute to the anisotropic
turbulent diffusivity (see, e.g., \citealp{kit-pip-rud,2003GApFD..97..249R,pi08Gafd}).
It is rather important for the dynamo wave propagation inside the
convection zone \citep{k02,kit:03}. 

The DNS dynamo experiments support the existence of the dynamo effects
induced by the large-scale current and global rotation \citep{2005AN....326..245S,2008A&A...491..353K}.
It was found that we have to account the complete expression of the
$\boldsymbol{\mathcal{E}}$ ( see, Eq. \ref{eq:emf}) to reproduce
the simulations of the global dynamo action \citep{2005AN....326..245S,2011A&A...533A.108S,2011MNRAS.418L.133M}
and evolution of the large-scale fields in the convective rotating
turbulent flows \citep{2009A&A...500..633K,bran2012AA}). The aim
of the DNS is to simulate the dynamo action in the cosmic bodies and
we are still on the way to reproduce the basic properties of the large-scale
dynamo for the Sun. It was shown that the mean-field MHD framework
is useful for the analysis of the results obtained in simulations
(see, also, \citealt{2011ApJ...735...46R}).

\section{The magnetic helicity issue}

The properties of the symmetry transformation of the $\boldsymbol{\mathcal{E}}$
suggest \citep{krarad80} that the $\alpha$ effect is pseudoscalar
(lacks the mirror symmetry) which is related to the kinetic helicity
of the small-scale flows, i.e., ${\displaystyle \alpha_{0}=-\frac{\tau_{c}}{3}\overline{\mathbf{u}\cdot\boldsymbol{\nabla\times}\mathbf{u}}}$.
\citet{pouquet-al:1975b} showed that the $\alpha$ effect is produced
not only by kinetic helicity but also by the current helicity, and
it is ${\displaystyle \alpha_{0}=-\frac{\tau_{c}}{3}\left(\overline{\mathbf{u}\cdot\boldsymbol{\nabla\times}\mathbf{u}}-\frac{\mathbf{\overline{b\cdot\boldsymbol{\nabla\times b}}}}{2\mu\overline{\rho}}\right)}$.
The latter effect can be interpreted as resistance of magnetic fields
against to twist by helical motions. It leads to the concept of the
catastrophic quenching of the $\alpha$ effect by the generated large-scale
magnetic field. It was found that ${\displaystyle \alpha_{0}\left(\overline{B}\right)=\frac{\alpha_{0}\left(0\right)}{1+R_{m}\left(\overline{B}/\overline{B}_{eq}\right)^{2}}}$
\citep{kle-rog99}. In case of $R_{m}\gg1$, the $\alpha$ effect
is quickly saturated for the large-scale magnetic field strength that
is much below the equipartition value ${\displaystyle \overline{B}_{eq}\sim\sqrt{\overline{\rho}\mu_{0}\overline{u^{2}}}}$.
The result was confirmed by the DNS\citep{oss2001}. The catastrophic
quenching (CQ) is related to the dynamical quenching of the $\alpha$
effect. It is based on conservation of the magnetic helicity, $\chi=\overline{\mathbf{a\cdot}\mathbf{b}}$
($\mathbf{a}$ is fluctuating part of the vector potential) and the
relation between the current and magnetic helicities $h_{\mathcal{C}}=\mathbf{\overline{b\cdot\boldsymbol{\nabla\times b}}\sim\chi}/\ell^{2}$,
which is valid for the isotropic turbulence\citep{moff:78}. The evolution
equation for $\chi$ can be obtained from equations that governs $\mathbf{a}$
and $\mathbf{b}$, it reads as follows \citep{kle-rog99,sub-bra:04}:
\begin{eqnarray}
\frac{\partial\overline{\chi}}{\partial t} & = & -2\left(\boldsymbol{\mathcal{E}}\cdot\overline{\bm{B}}\right)-\frac{\overline{\chi}}{R_{m}\tau_{c}}-\boldsymbol{\nabla}\cdot\boldsymbol{\boldsymbol{\mathcal{F}}}^{\chi}-\eta\overline{\mathbf{B}}\cdot\mathbf{\overline{J}},\label{eq:hel}
\end{eqnarray}
where we introduce the helicity fluxes $\boldsymbol{\boldsymbol{\mathcal{F}}}^{\chi}=\mathbf{\overline{a\times u}}\times\mathbf{B}-\mathbf{\overline{a\times(u\times b)}}$.
The helicity fluxes are capable to alleviate the catastrophic quenching
(CQ). The first example was given for the galactic dynamo model \citep{2000A&A...361L...5K}.
The calculations and the DNS shows the existence of the of the several
kind of the helicity fluxes. Part of them have the turbulent origin,
e.g., the fluxes due to the anisotropy of the turbulent flows (Kleeorin
\& Rogachevskii 1999), the fluxes due to the large-scale shear \citep{vish-ch:01}
and the diffusive fluxes \citep{mitra10}. Another kind of the helicity
fluxes, which are not mentioned in the Eq.(\ref{eq:hel}), are related
to the large-scale flows, e.g., meridional circulation and outflows
due to the solar wind \citep{mitra11}. Generally, it was found that
the diffusive fluxes, which are $\sim\eta_{\chi}\boldsymbol{\nabla}\chi$,
where $\eta_{\chi}$ is the turbulent diffusivity of the magnetic
helicity, work robustly in the mean-field dynamo models but it requires
$\eta_{\chi}>\eta_{T}$ to reach $\left|\overline{B}\right|\ge0.1\overline{B}_{eq}$. 

Another possibility to alleviate the catastrophic quenching is related
with the non-local formulation of the mean-electromotive force\citep{bs02,2008A&A...482..739B}.
\citet{2011AstL...37..286K} found that the nonlocal $\alpha$ effect and the diamagnetic pumping
can alleviate the catastrophic quenching. The results by \citet{2007NJPh....9..305B}
show that the result can depend on the model design. Nonlocal formulation
of the mean-field MHD concept suggests a possibility to solve the
problem related to the dynamo period in the mean-field
models. \cite{2012AN....333...71R} showed that
corrections for nonlocal effect in the mean-electromotive
force can be taken into account with the partial equation like $\left(1+\tau_{c}\partial_{t}+\ell^{2}\Delta\right)\boldsymbol{\mathcal{E}}_{i}=\boldsymbol{\mathcal{E}}_{i}^{(0)}$,
where$\boldsymbol{\mathcal{E}}_{i}^{(0)}$ is the local version of
the mean-electromotive force given by Eq.(\ref{eq:emf}).

We have to notice, the solar dynamo is an open system, where the
large-scale magnetic fields escape from the dynamo region to the outer
atmosphere. 
For the vacuum boundary conditions, which are widely used in
the solar dynamo models, the magnetic field escapes freely from the
solar convection zone, and nothing prevents the magnetic
helicity accompanying the large-scale magnetic field to escape the
dynamo region. Thus, the magnetic helicity conservation should not
pose an issue for the solar type dynamos.  
Recently, \citet{2012ApJ...748...51H} revisited the CQ concept and
showed that for the shearing dynamos the Eq.(\ref{eq:hel}) produces
the nonphysical fluxes of the magnetic helicity over the spatial scales.
\citet{2012ApJ...748...51H} suggested to cure the situation starting
from the global conservation law for the magnetic helicity,
\begin{equation}
\frac{d}{dt}\int\left\{ \overline{\mathbf{a\cdot}\mathbf{b}}+\overline{\mathbf{A}}\cdot\overline{\mathbf{B}}\right\} dV=-\eta\int\left\{ \overline{\mathbf{B}}\cdot\mathbf{\overline{J}}+\overline{\mathbf{b\cdot j}}\right\} dV-\int\boldsymbol{\nabla\cdot}\boldsymbol{\boldsymbol{\mathcal{F}}}^{\chi}dV\label{eq:int-cons}
\end{equation}
where integration is done over the volume that comprises the ensemble
of the small-scale fields. We assume that $\boldsymbol{\boldsymbol{\mathcal{F}}}^{\chi}$
is the diffusive flux of the total helicity which is resulted from
the turbulent motions. Ignoring the effect of the meridional circulation
we write the local version of the Eq.(\ref{eq:int-cons}) as follows
\citep{2012ApJ...748...51H}:

\begin{equation}
\partial_{t}\overline{\mathbf{a\cdot}\mathbf{b}}=-\partial_{t}\left(\overline{\mathbf{A}}\cdot\overline{\mathbf{B}}\right)-\frac{\overline{\chi}}{R_{m}\tau_{c}}-\eta\overline{\mathbf{B}}\cdot\mathbf{\overline{J}}-\boldsymbol{\nabla\cdot}\boldsymbol{\boldsymbol{\mathcal{F}}}^{\chi}.\label{eq:helcon2}
\end{equation}
Note, that the large-scale helicity is govern by:
\begin{eqnarray}
\partial_{t}\left(\overline{\mathbf{A}}\cdot\overline{\mathbf{B}}\right) & = & 2\boldsymbol{\mathcal{E}}\cdot\overline{\mathbf{B}}+\boldsymbol{\nabla}\cdot\left(\left(\boldsymbol{\mathcal{E}}\times\overline{\mathbf{A}}\right)-\mathbf{\overline{A}}\times\left(\overline{\mathbf{U}}\times\overline{\mathbf{B}}\right)\right).\label{AB}
\end{eqnarray}
Therefore, Eqs.(\ref{eq:hel}) and (\ref{eq:helcon2}) differ by the
second part of Eq.(\ref{AB}). \citet{2012ApJ...748...51H} found
that the $\boldsymbol{\nabla}\cdot\left(\boldsymbol{\mathcal{E}}\times\overline{\mathbf{A}}\right)$
cures the problem the nonphysical fluxes of the magnetic helicity in
shearing systems. Another term $\boldsymbol{\nabla}\cdot\left(\mathbf{\overline{A}}\times\left(\overline{\mathbf{U}}\times\overline{\mathbf{B}}\right)\right)$
contains the transport of the large-scale magnetic helicity by the
large-scale flow. It was found that the dynamos with the dynamical
quenching govern by the Eq.(\ref{eq:helcon2}) does not suffer from
the catastrophic quenching issue.

\section{The dynamo shaped by the subsurface shear layer}

Most of the solar dynamo models suggest that the toroidal magnetic
field that emerges on the surface and forms sunspots is generated
near the bottom of the convection zone, in the tachocline or just
beneath it in a convection overshoot layer, (see, e.g., \citealp{1995A&A...296..557R,2002A&A...390..673B,2006ApJ...647..662R,2008A&A...485..267G,2009RMxAC..36..252G}).
However, an attention was drawn to a number of theoretical and observational
problems concerning the deep-seated dynamo models \citep{2005ApJ...625..539B,2006ASPC..354..121B}.
We proposed (see \citealp{pk11apjl,pk11}) a solar dynamo model distributed
in the bulk of the convection zone with toroidal magnetic-field flux
concentrated in a near-surface layer. In this section I would like
to illustrate the theoretical profiles for the principal parts of
the mean-electromotive force for the linear and nonlinear case in
the mean-field large-scale dynamo and compare them with the results
of the DNS. 

\begin{figure}
\noindent \begin{centering}
\includegraphics[width=0.42\textwidth]{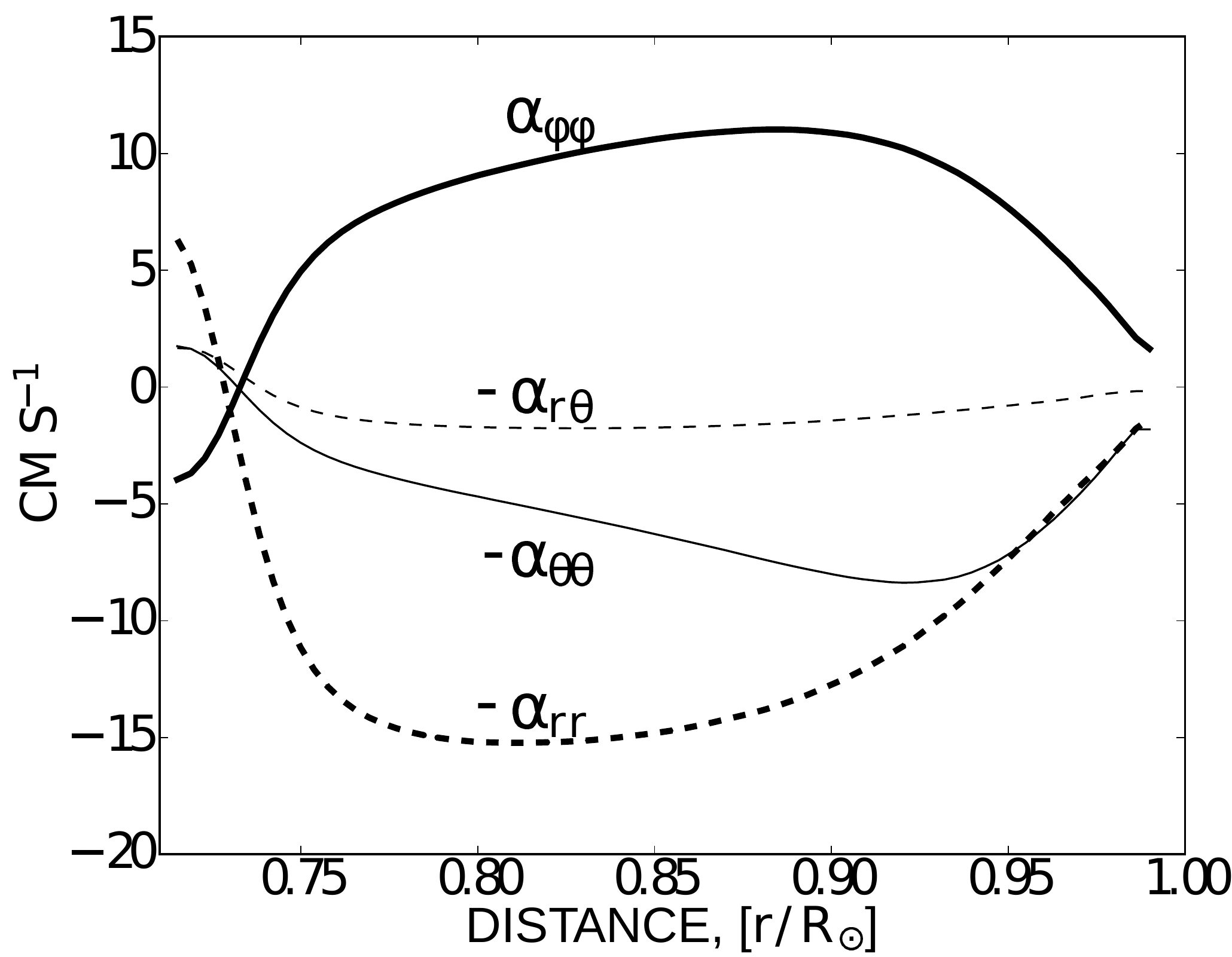}\includegraphics[width=0.45\textwidth]{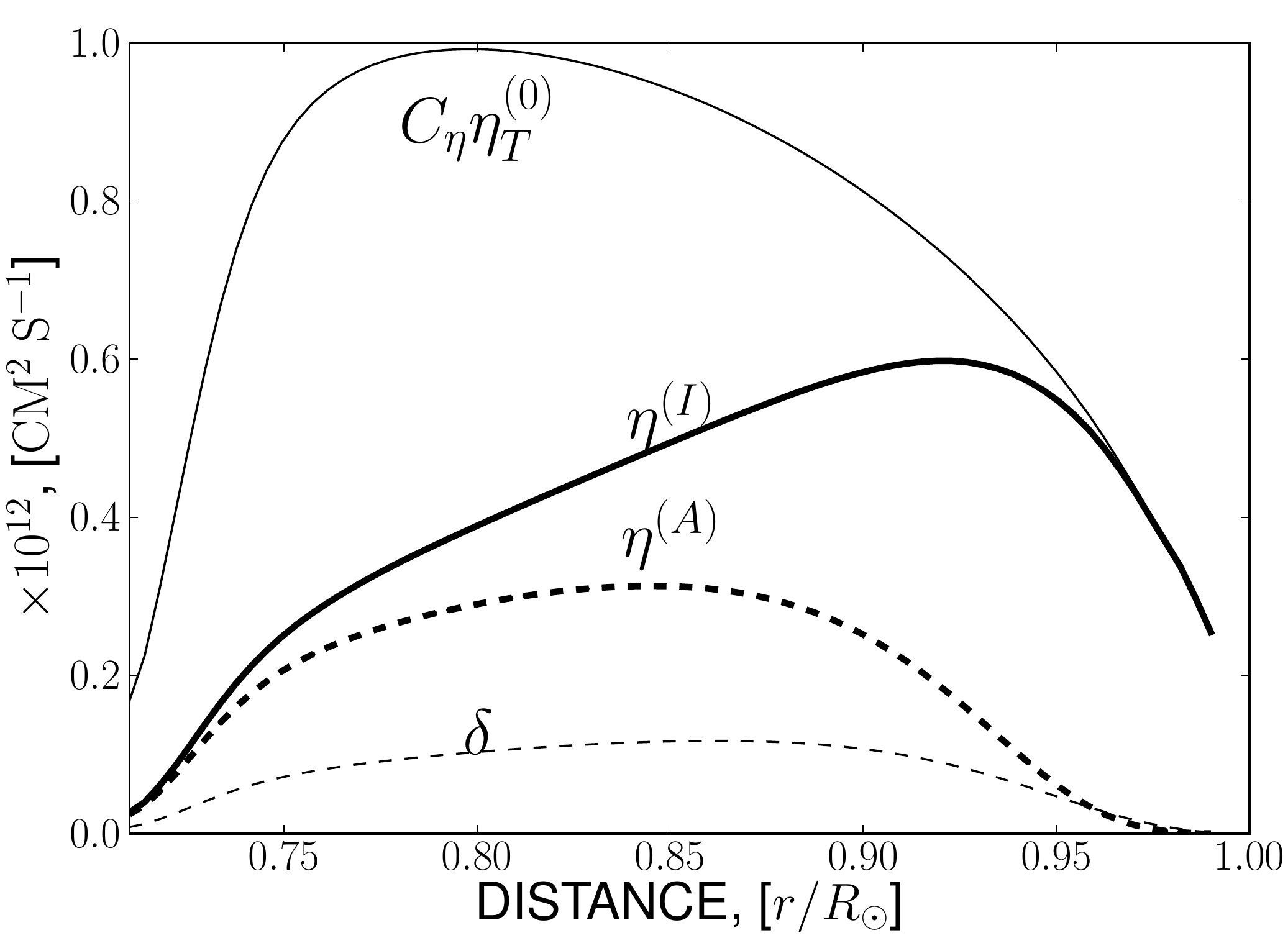}
\par\end{centering}
\caption{Left, the profiles of the $\alpha$ effects components for the $\theta=45^{\circ}$.
Right, the profiles of the background turbulent diffusivity $C_{\eta}\eta_{T}$,
the isotropic, $\eta^{(I)}$, and anisotropic, $\eta^{(A)}$, parts
of the magnetic diffusivity and $\Omega\times J$ effect (also known
as $\delta$ effect \citep{rad69,stix76a}\label{fig:profiles}.}
\end{figure}

We study the mean-field dynamo equation Eq.(\ref{eq:emf}) for the
axisymmetric magnetic field $\overline{\bm{B}}=\bm{e}_{\phi}B+\nabla\times\left(A\bm{e}_{\phi}/\left(r\sin\theta\right)\right),$
where $\theta$ is the polar angle. The expression for the mean electromotive
force $\boldsymbol{\mathcal{E}}$ is given by \citet{pi08Gafd}: 
\begin{equation}
\mathcal{E}_{i}=\left(\alpha_{ij}+\gamma_{ij}^{(\Lambda)}\right)\overline{B}_{j}-\eta_{ijk}\nabla_{j}\overline{B}_{k}.\label{eq:EMF-1}
\end{equation}
The tensor $\alpha_{ij}$ represents the $\alpha$-effect. It includes
hydrodynamic and magnetic helicity contributions, 
\begin{eqnarray}
\alpha_{ij} & = & C_{\alpha}\sin^{2}\theta\alpha_{ij}^{(H)}+\alpha_{ij}^{(M)},\label{alp2d}\\
\alpha_{ij}^{(H)} & = & \delta_{ij}\left\{ 3\eta_{T}\left(f_{10}^{(a)}\left(\bm{e}\cdot\boldsymbol{\Lambda}^{(\rho)}\right)+f_{11}^{(a)}\left(\bm{e}\cdot\boldsymbol{\Lambda}^{(u)}\right)\right)\right\} +\\
 & + & e_{i}e_{j}\left\{ 3\eta_{T}\left(f_{5}^{(a)}\left(\bm{e}\cdot\boldsymbol{\Lambda}^{(\rho)}\right)+f_{4}^{(a)}\left(\bm{e}\cdot\boldsymbol{\Lambda}^{(u)}\right)\right)\right\} +\nonumber \\
 &  & 3\eta_{T}\left\{ \left(e_{i}\Lambda_{j}^{(\rho)}+e_{j}\Lambda_{i}^{(\rho)}\right)f_{6}^{(a)}+\left(e_{i}\Lambda_{j}^{(u)}+e_{j}\Lambda_{i}^{(u)}\right)f_{8}^{(a)}\right\} ,\nonumber 
\end{eqnarray}
where the hydrodynamic part of the $\alpha$-effect is defined by
$\alpha_{ij}^{(H)}$, $\mathbf{\Lambda}^{(\rho)}=\boldsymbol{\nabla}\log\overline{\rho}$
quantifies the density stratification, $\mathbf{\Lambda}^{(u)}=C_{v}\boldsymbol{\nabla}\log\left(\eta_{T}^{(0)}\right)$
quantifies the turbulent diffusivity variation, and $\bm{e}=\boldsymbol{\Omega}/\left|\boldsymbol{\Omega}\right|$
is a unit vector along the axis of rotation. The turbulent pumping,
$\gamma_{ij}^{(\Lambda)}$, depends on mean density and turbulent
diffusivity stratification, and on the Coriolis number $\Omega^{*}=2\tau_{c}\Omega_{0}$
where $\tau_{c}$ is the typical convective turnover time and $\Omega_{0}$
is the global angular velocity. We introduce the parameter $C_{v}$
for the $\mathbf{\Lambda}^{(u)}$ to take into account the results
from the DNS \citep{oss2001,2009A&A...500..633K} which show that
the $\alpha$ effect is saturated to the bottom of the convection
zone. We will show the profiles below. Following the results of \citet{pi08Gafd},
$\gamma_{ij}^{(\Lambda)}$ is expressed as follows: 
\begin{eqnarray}
\gamma_{ij}^{(\Lambda)} & = & 3\eta_{T}\left\{ f_{3}^{(a)}\Lambda_{n}^{(\rho)}+f_{1}^{(a)}\left(\bm{e}\cdot\boldsymbol{\Lambda}^{(\rho)}\right)e_{n}\right\} \varepsilon_{inj}-3\eta_{T}f_{1}^{(a)}e_{j}\varepsilon_{inm}e_{n}\Lambda_{m}^{(\rho)},\label{eq:pump}\\
 & - & 3\eta_{T}\left(\varepsilon-1\right)\left\{ f_{2}^{(a)}\Lambda_{n}^{(u)}+f_{1}^{(a)}\left(\bm{e}\cdot\boldsymbol{\Lambda}^{(u)}\right)e_{n}\right\} \varepsilon_{inj}.\nonumber 
\end{eqnarray}
The effect of turbulent diffusivity, which is anisotropic due to the
Coriolis force, is given by: 
\begin{equation}
\eta_{ijk}=3\eta_{T}\left\{ \left(2f_{1}^{(a)}-f_{2}^{(d)}\right)\varepsilon_{ijk}-2f_{1}^{(a)}e_{i}e_{n}\varepsilon_{njk}+\varepsilon C_{\omega}f_{4}^{(d)}e_{j}\delta_{ik}\right\} .\label{eq:diff}
\end{equation}
The functions $f_{\{1-11\}}^{(a,d)}$in Eqs(\ref{alp2d},\ref{eq:pump},\ref{eq:diff})
depend on the Coriolis number. They can be found in \citet{pi08Gafd}
(see also, \citealp{pk11,ps11}). In the model, the parameter $\varepsilon={\displaystyle \frac{\overline{\bm{b}^{2}}}{\mu_{0}\overline{\rho}\overline{\bm{u}^{2}}}}$,
which measures the ratio between magnetic and kinetic energies of
the fluctuations in the background turbulence, is assumed to be equal
to 1. In our models we use the solar convection zone model computed
by \citet{stix:02}. The mixing-length is defined as $\ell=\alpha_{{\rm MLT}}\left|\Lambda^{(p)}\right|^{-1}$,
where $\Lambda{}^{(p)}=\boldsymbol{\nabla}\log\overline{p}\,$ quantifies
the pressure variation, and $\alpha_{{\rm MLT}}=2$. The turbulent
diffusivity is parameterized in the form, $\eta_{T}=C_{\eta}\eta_{T}^{(0)}$,
where $\eta_{T}^{(0)}={\displaystyle \frac{u'^{2}\tau_{c}}{3f\left(r\right)}}$
is the characteristic mixing-length turbulent diffusivity, $\ell$
is the typical correlation length of the turbulence, $C_{\eta}$ is
a constant to control the efficiency of large-scale magnetic field
dragging by the turbulent flow. Also, we modify the mixing-length
turbulent diffusivity by factor $f_{s}(r)=1+\exp\left(100\left(r_{ov}-r\right)\right)$,
$r_{ov}=0.72R_{\odot}$ to get the saturation of the turbulent parameters
to the bottom of the convection zone.  The latter is suggested by
the DNS. The results do not change very much if we apply $\mathbf{\Lambda}^{(u)}=C_{v}\boldsymbol{\nabla}\log\left(\eta_{T}^{(0)}\right)$
with $C_{v}\le0.5$. For the greater $C_{v}$ we get the steady non-oscillating
dynamo concentrated to the bottom of the convection zone. I would like
to stress that the purpose to introduce the additional parameters
like $C_{v}=0.5$ and $f_{s}(r)$ is to get the distribution of the
$\alpha$ effect closer to the result obtained in the DNS. The bottom
of the integration domain is $r_{b}=0.71R_{\odot}$ and the top of
the integration domain is $r_{e}=0.99R_{\odot}$. The choice of parameters
in the dynamo is justified by our previous studies \citep{2009A&A...493..819P,pk11mf},
where it was shown that solar-types dynamos can be obtained for $C_{\alpha}/C_{\omega}>2$.
In those papers we find the approximate threshold to be $C_{\alpha}\approx0.03$
for a given diffusivity dilution factor of $C_{\eta}=0.1$. Figure
\ref{fig:profiles} shows the radial profiles for the principal components
of the mean electromotive force, which are essential for our model.
They are in the qualitative agreement with the results of the DNS
obtained by \citet{oss2001} and \citet{2009A&A...500..633K}.
The contribution of small-scale magnetic helicity $\overline{\chi}=\overline{\bm{a\cdot}\bm{b}}$
to the $\alpha$-effect is defined as 
\begin{equation}
\alpha_{ij}^{(M)}=2f_{2}^{(a)}\delta_{ij}\frac{\overline{\chi}\tau_{c}}{\mu_{0}\overline{\rho}\ell^{2}}-2f_{1}^{(a)}e_{i}e_{j}\frac{\overline{\chi}\tau_{c}}{\mu_{0}\overline{\rho}\ell^{2}}.\label{alpM}
\end{equation}
The nonlinear feedback of the large-scale magnetic field to the $\alpha$-effect
is described by a dynamical quenching due to the constraint of magnetic
helicity conservation given by Eq.(\ref{eq:helcon2}). For the illustration
purpose we use the realistic value for the magnetic Reynolds number
$R_{m}=10^{6}$. We matched the potential field outside and the perfect
conductivity at the bottom boundary with the standard boundary conditions.
For the magnetic helicity the number of the possibilities can be used
\citep{guero10,mitra10}. 
We employ
$\nabla_{r}\left(\bar{\chi}+\overline{\mathbf{A}}\cdot\overline{\mathbf{B}}\right)=0$
at the top of the domain and $\bar{\chi}=0$ at the bottom of the
convection zone to show that for the $\alpha$-quenching formalism
which is based on the Eq.(\ref{eq:helcon2}), the boundary conditions
determine the dynamics of the dynamo wave at the near surface layer.
To evolve the Eq.(\ref{eq:helcon2}) we have to define the large-scale
vector potential for each time-step. For the axisymmetric large-scale
magnetic fields where the vector-potential is 
\begin{equation}
\overline{\mathbf{A}}=\mathbf{e}_{\phi}P+\mathbf{r}T=\frac{R^{2}\mathbf{e}_{\phi}}{r\sin\theta}A+r\mathbf{e}_{r}T.\label{eq:potent}
\end{equation}
The toroidal part of the vector potential is governed by the dynamo
equations. The poloidal part of the vector potential can be restored
from equation $\boldsymbol{\nabla}\times\left(\mathbf{r}T\right)=\mathbf{e}_{\phi}B$. 

\begin{figure}
\noindent \begin{centering}
\includegraphics[width=0.8\textwidth]{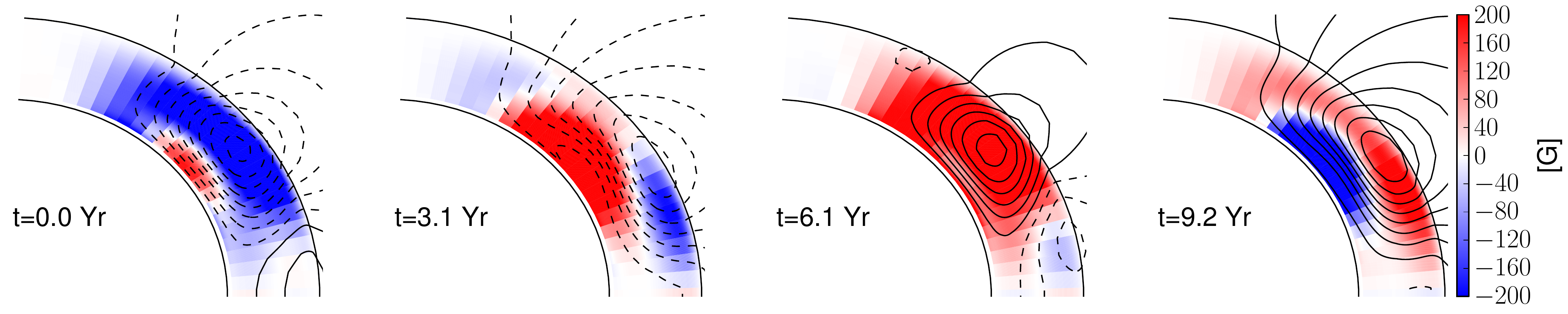}
\par\end{centering}

\noindent \begin{centering}
\includegraphics[width=0.8\textwidth]{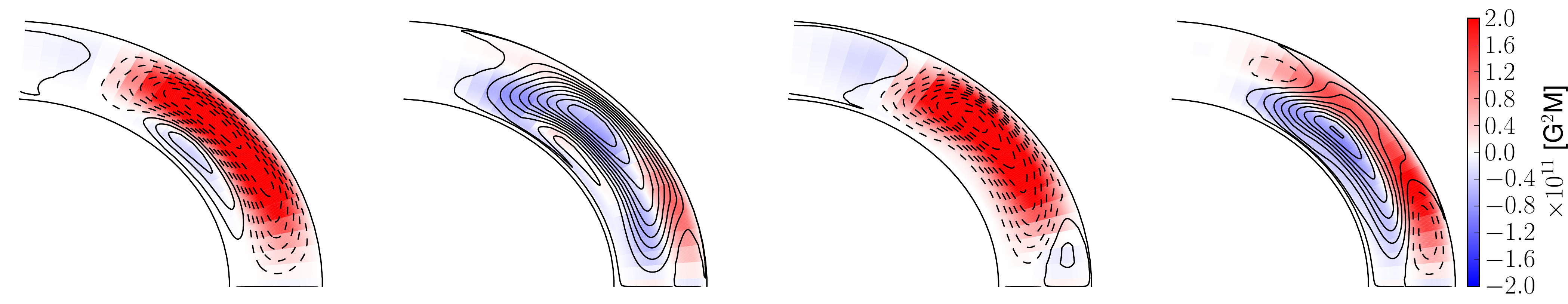}
\par\end{centering}
\caption{Snapshots of the magnetic field and helicity evolution inside the
North segment of the convection zone. Top panel panel shows the field
lines of the poloidal component of the mean magnetic field and the
toroidal magnetic field (varies $\pm700$G) by gray scale density
plot. The bottom panel shows the large-scale (density plot) and small-scale
magnetic helicity (contours) distributions.
\label{fig:Snapshots}}
\end{figure}
Figure \ref{fig:Snapshots} shows the snapshots of the magnetic field
and magnetic helicity (large- and small-scale) evolution in the North
segment of the solar convection zone. The Figure shows the drift of
the dynamo waves which are related to the large-scale toroidal and
poloidal fields towards the equator and towards the pole, respectively.
The distributions of the large- and small-scale magnetic helicities
show one to one correspondence in sign. This is in agreement with
Eq.(\ref{eq:helcon2}). It is seen that the negative sign of the magnetic
helicity follows to the dynamo wave of the toroidal magnetic field.
This can be related to the so-called ``current helicity hemisperic
rule'' which is suggested by the observations \citep{see1990SoPh,zetal10}.
The origin of the helicity rule has been extensively studied in the
dynamo theory \citep{pevt99ASPC,choud2004ApJ,kps:06,soka06,pevts2007ASPC,pk11,2012ApJ...751...47Z}. 

\begin{figure}
\noindent \begin{centering}
\includegraphics[width=0.8\textwidth,height=0.14\textheight]{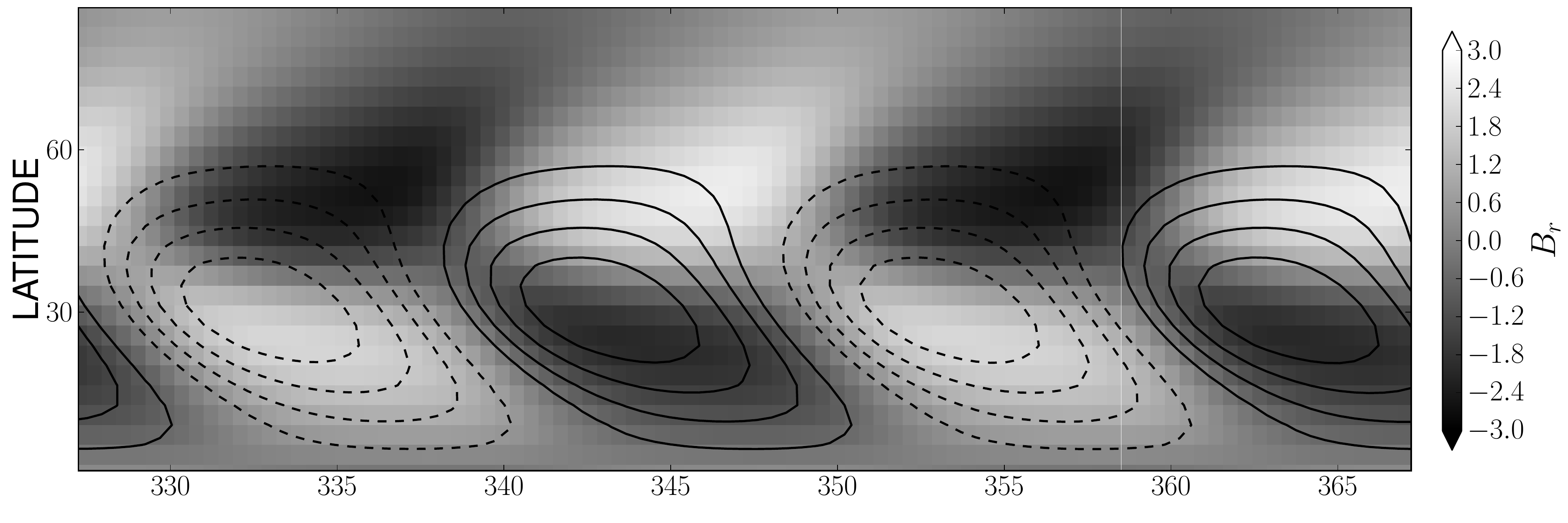}
\par\end{centering}

\noindent \begin{centering}
\includegraphics[width=0.8\textwidth,height=0.14\textheight]{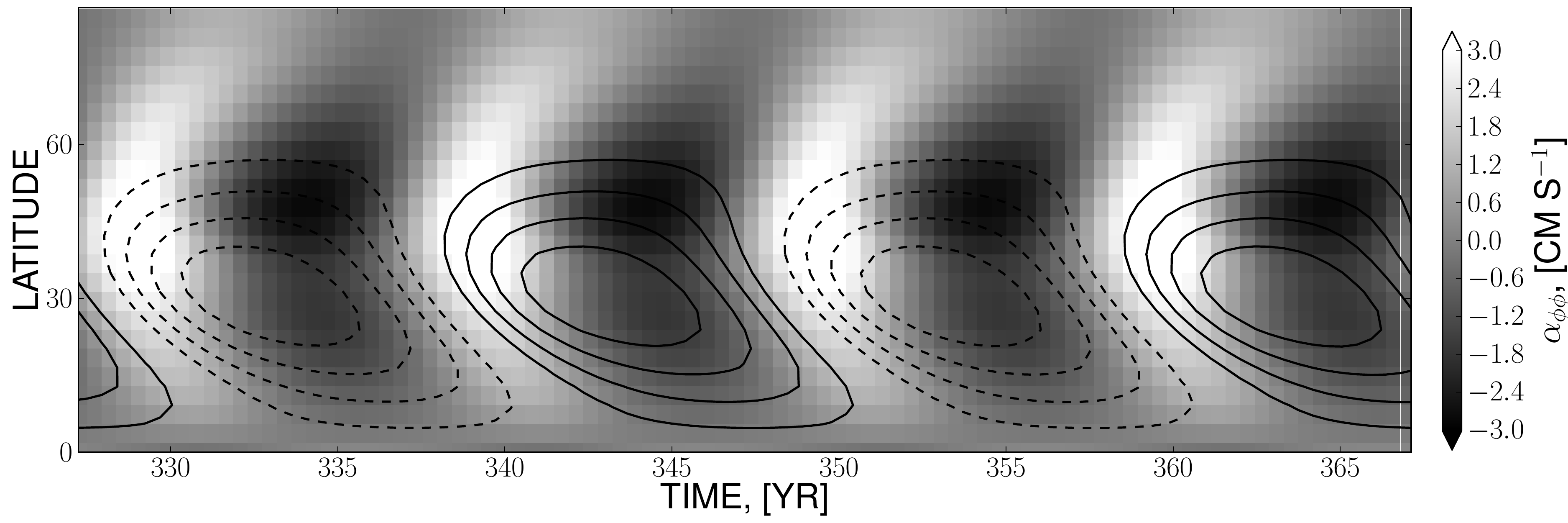}
\par\end{centering}

\caption{Top, the time - latitude variations of the toroidal field near the
surface, $r=0.95R_{\odot}$, (contours $\pm300$G) and the radial
magnetic field at the surface (density plot). Bottom, the same for
the the toroidal field and the $\alpha$ effect ($\alpha_{\phi\phi}$
component)(density plot).\label{fig:time-lat}}
\end{figure}

\begin{figure}
\noindent \begin{centering}
\includegraphics[width=0.38\textwidth]{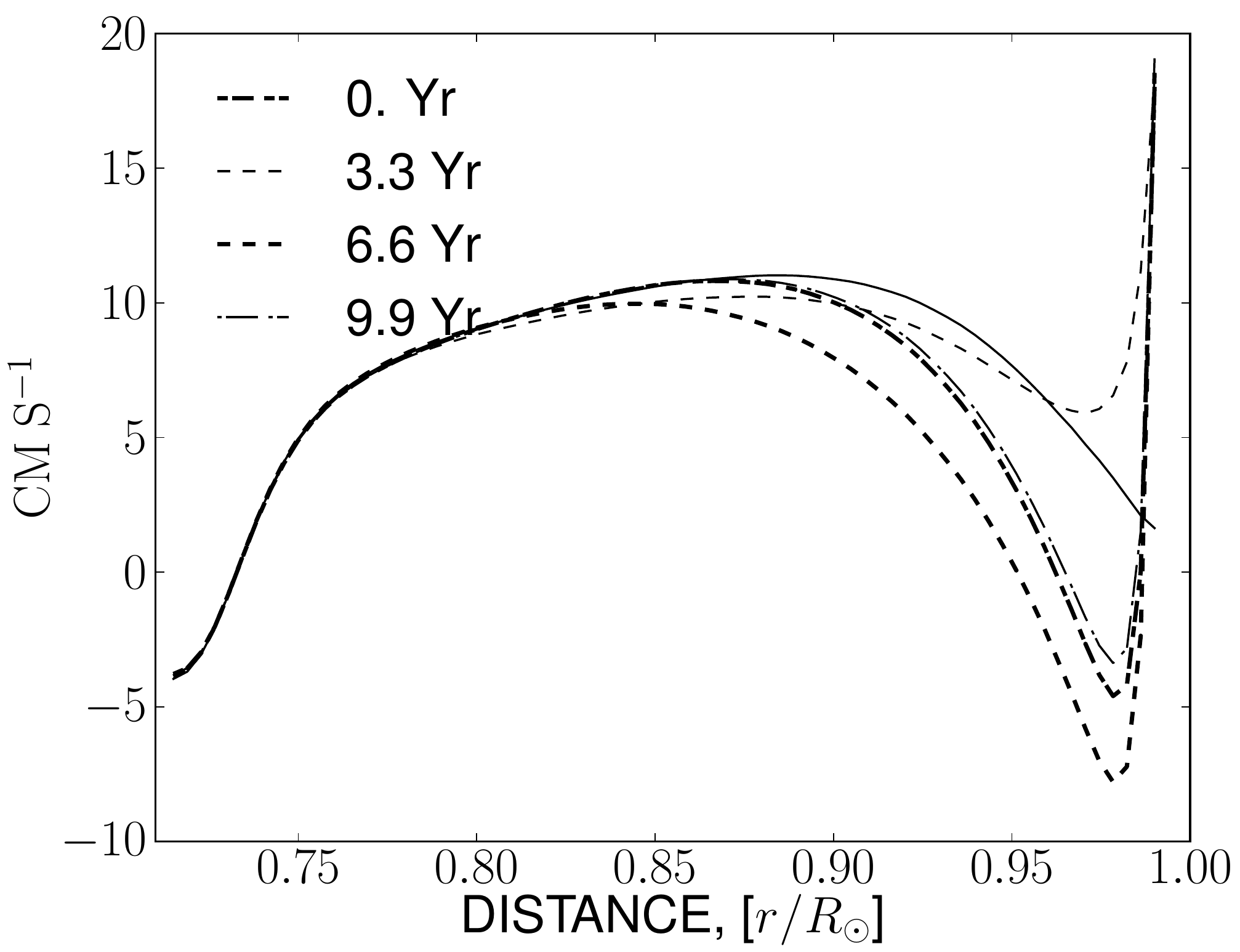}\includegraphics[width=0.4\textwidth]{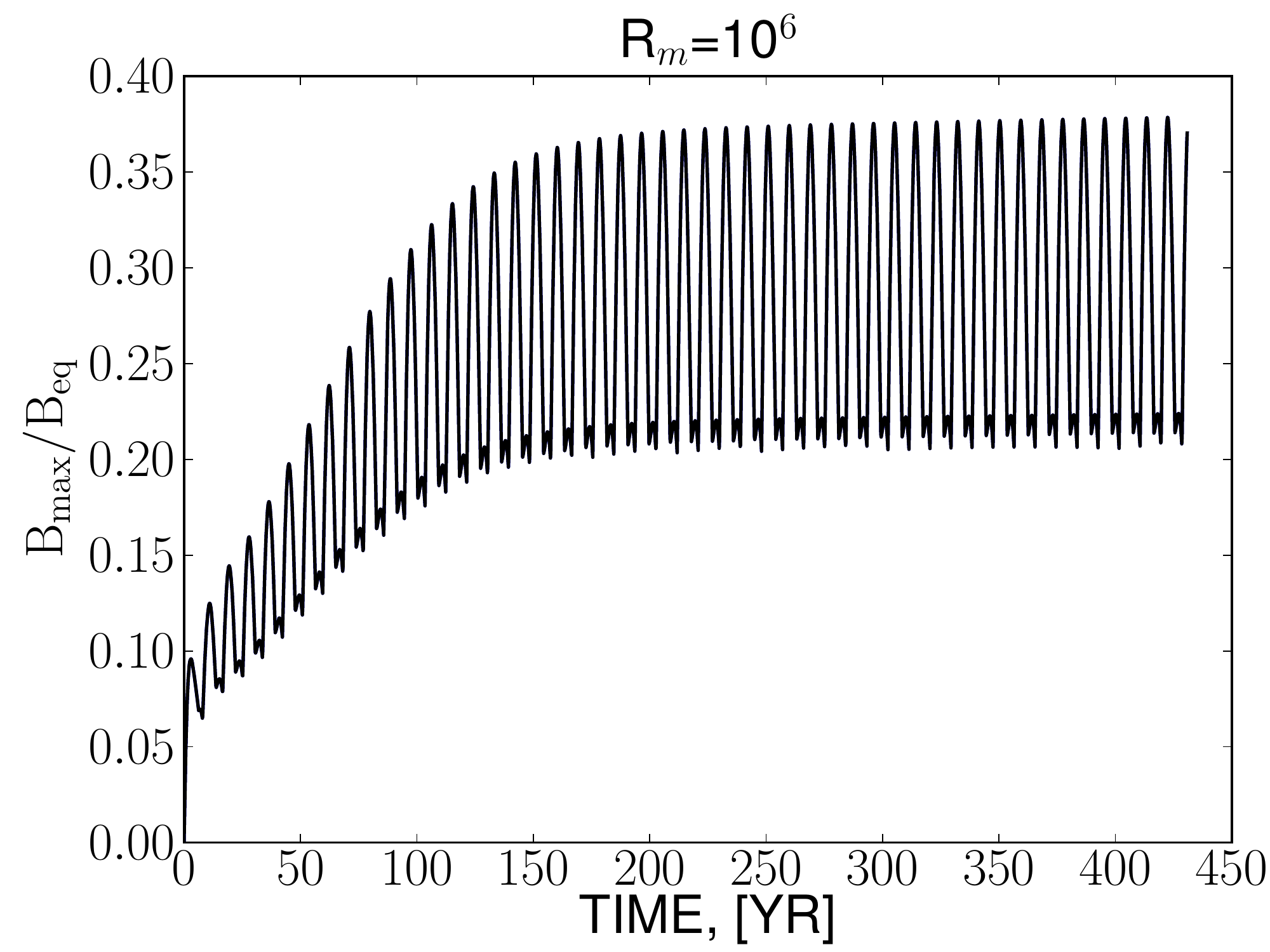}
\par\end{centering}

\caption{Left  panel shows variations of the $\alpha$ effect ($\alpha_{\phi\phi}$
component) profiles with the cycle at colatitude $\theta=45^{\circ}$,
the kinetic part of the $\alpha$ effect is shown by the solid line.
Right panel shows the ratio of the maximum of the toroidal field strength
and the equipartition value with time. \label{fig:helvar}.}
\end{figure}

The time-latitude diagrams for the dynamo model are shown in Figure
\ref{fig:time-lat}. The results are in qualitative agreement with observations.
We show the dynamical $\alpha$ effect as well. The model shows that
with the given boundary conditions the $\alpha$ effect increases
and has positive maximum at the growing phase of the cycle and it decreases,
having the negative minimum at the decaying phase of the cycle. The
variations of the radial profiles for the $\alpha$ effect and the
small-scale-magnetic helicity are shown in Figure \ref{fig:helvar}.
The saddle in the the $\alpha$ effect profile is resulted from the
given boundary conditions and distribution of the large-scale magnetic
field near the top of the solar convection zone. The latter is determined
by the dynamo boundary conditions and by the subsurface shear layer.
It was found that the saddle disappears in case of $\nabla_{r}\bar{\chi}=0$
at the top. The highly non-uniform radial distribution of the $\alpha$
effect (at least for the near-equatorial latitudes) was found in the
global LES (large-scale eddy) simulation of the dynamo action in the
recent paper by \citet{2011ApJ...735...46R}. 

The results obtained with the given dynamo model can be summarized
as follows. In the model the dynamo wave propagates through the convection
zone and shaped by the subsurface shear layer. The model incorporates
the fairly complete expressions for the mean-electromotive force,
including the anisotropic turbulent pumping and magnetic diffusivity,
the turbulent generation of the poloidal and the toroidal fields due
to $\alpha^{2}$ and $\Omega\times J$ effects and the dynamical $\alpha$
quenching due to magnetic helicity.
We demonstrate that the conservation of the total helicity given in the form of Eq.(\ref{eq:helcon2})
alleviates the catastrophic quenching (see Figure \ref{fig:helvar})
for the solar type dynamos that operates on the base of the $\alpha\Omega$
mechanism. The further development can be related to the nonlocal framework for the mean-electromotive
force.


\section{Conclusions }

Summarizing the main topics of this review, which I also consider
as a principal advances in the mean-field dynamo theory, I conclude
as follows. Firstly, the knowledge of the mean-field coefficients
is decisive in order to analyze and model dynamo action in a variety
astrophysical bodies. It was found that agreement between the direct
numerical simulation and mean-field models is good only if a large
number of mean-field coefficients are involved, which contribute to
$\alpha$, $\gamma$, $\eta$, $\kappa$ (see, Eq.\ref{eq:emf}).
One obvious reason for this is that the spatial scale-separation is
not valid for many astrophysical cases. Secondly, 
the issues related to the catastrophic quenching of the dynamo can
be alleviated by different ways. However, the approach,
which is based on conservation of the total helicity, see Eq.(\ref{eq:helcon2}),
suggests just a more than the way around the catastrophic quenching
issue. It was proved that the magnetic helicity conservation does not
pose an issue for the solar type dynamos. The new formalism
 can be used to study the origin of the different kind of
anomalies in solar magnetic activity, which are observed and related
to evolution of the large-scale toroidal field, the surface distribution
of the magnetic helicity at different phase of the solar cycle etc.
These points have not been discussed in this review and remain an
open field for the future work. 

{\bf Acknowledgments}
\medskip
I thank for the support the RFBR grants
12-02-00170-a, 10-02-00148-a and 10-02-00960,  the support of the  Integration Project of SB RAS
N 34, and  support of the state contracts 02.740.11.0576, 16.518.11.7065 of the Ministry
 of Education and Science of Russian Federation.


\end{document}